\documentclass[%
superscriptaddress,
nofootinbib,
twocolumn]{revtex4-2}

\usepackage{amssymb, amsmath, bm, dcolumn, epsf, graphicx, latexsym, slashed, simplewick}
\usepackage[utf8]{inputenc}

\usepackage{graphicx}
\usepackage{dcolumn}
\usepackage{bm}
\usepackage{xcolor}
\usepackage{bbold}
\usepackage{soul}
\usepackage[usestackEOL]{stackengine}
\usepackage{color}

\usepackage{amsmath}
\usepackage{amssymb}
\usepackage[normalem]{ulem}

\begin{document}

\preprint{APS/123-QED}

\title{Decoding Dark Matter Substructure without Supervision}

\author{Stephon~Alexander}
\affiliation{Brown Theoretical Physics Center, Providence, RI 02912, USA}
\affiliation{Department of Physics, Brown University, Providence, RI 02912, USA}

\author{Sergei~Gleyzer}
\affiliation{Department of Physics \& Astronomy, University of Alabama, Tuscaloosa, AL 35401, USA}

\author{Hanna~Parul}
\affiliation{Department of Physics \& Astronomy, University of Alabama, Tuscaloosa, AL 35401, USA}

\author{Pranath~Reddy}
\affiliation{Birla Institute of Technology \& Science, Pilani - Hyderabad Campus, Telangana, India}

\author{Michael~W.~Toomey}
\affiliation{Brown Theoretical Physics Center, Providence, RI 02912, USA}
\affiliation{Department of Physics, Brown University, Providence, RI 02912, USA}

\author{Emanuele~Usai}
\affiliation{Department of Physics, Brown University, Providence, RI 02912, USA}

\author{Ryker~Von~Klar}
\affiliation{Department of Physics \& Astronomy, University of Alabama, Tuscaloosa, AL 35401, USA}

\date{\today}

\begin{abstract}
The identity of dark matter remains one  of the most pressing questions in physics today. While many promising dark matter candidates have been put forth over the last half-century, to date the true identity of dark matter remains elusive. While it is possible that one of the many proposed candidates may turn out to be dark matter, it is at least equally likely that the correct physical description has yet to be proposed. To address this challenge, novel applications of machine learning can help physicists gain insight into the dark sector from a {\it theory agnostic} perspective. In this work we demonstrate the use of unsupervised machine learning techniques to infer the presence of substructure in dark matter halos using galaxy-galaxy strong lensing simulations.\\
\end{abstract}

\maketitle


\section{Introduction}

Since the discovery of dark matter, physicists have been searching the entirety of cosmic history for fingerprints that might reveal its identity, from experiments at colliders to observations of the cosmic microwave background. Most dark matter models assume that the dark sector interacts, typically only very weakly, with the Standard Model - e.g. WIMPs \cite{Steigman:1984ac} and axions \cite{Preskill:1982cy,Abbott:1982af,Dine:1982ah}. However, direct and indirect detection searches \cite{Drukier:1986tm,Goodman:1984dc,Akerib:2016vxi,Cui:2017nnn,Aprile:2018dbl,2020arXiv200304545F,2015JCAP...09..008F,2015PhRvD..91h3535G,2018ApJ...853..154A,2017PhRvD..95h2001A,2020Galax...8...25R,2016JCAP...02..039M,2017arXiv170508103I,2015arXiv150304858T,2018PhRvL.120o1301D, 2015ARNPS..65..485G,2020arXiv200610735K,2020arXiv200612488B}, including searches at colliders \cite{Aaboud:2019yqu,2017JHEP...10..073S} have not yielded a discovery. To date the only evidence for dark matter comes from its gravitational interactions \cite{planck,clust,wlens,Clowe:2003tk}. This makes a strong case of exploring new avenues to identify dark matter via its gravitational fingerprints.

A promising path to identify the nature of dark matter is to study substructure in dark matter halos. Variation between model predictions on {\it subgalactic} scales will allow current and future observational programs to begin to constrain potential dark matter candidates \cite{Buckley:2017ijx, Drlica-Wagner:2019xan, Simon:2019kmm}. While it is possible to study larger substructures such as ultra-faint dwarf galaxies (for example  \cite{Drlica-Wagner:2015ufc}), subhalos on smaller scales have suppressed star formation, making manifest the need for a gravitational probe. Promising directions to identify substructure gravitationally include tidal streams \cite{Ngan:2013oga, 2016ApJ...820...45C, 2016PhRvL.116l1301B, 2016MNRAS.463..102E} and astrometric observations \cite{Mishra-Sharma:2020ynk, VanTilburg:2018ykj, Feldmann:2013hqa, 2016arXiv160805624S}. Another avenue to consider is strong gravitational lensing which has seen encouraging results in detecting the existence of substructure from strongly lensed quasars \cite{sub1,sub2,sub3}, high resolutions observations with the Atacama Large Millimeter/submillimeter Array \cite{alma} and, extended lensing images \cite{2010MNRAS.408.1969V,2012Natur.481..341V,2014MNRAS.442.2017V,Ritondale:2018cvp,veg,koop,veko}.

Bayesian likelihood analyses have been proposed as a first approach to identifying dark matter by determining if a given model is consistent with a set of lensing images – analyses of this nature have been performed in the context of particle dark matter substructure \cite{pcat,subs}. These approaches can be limited by significant computational cost, while approaches that rely on machine learning algorithms, can produce nearly instantaneous results at inference stage.

Machine learning, and particularly deep learning, methods have wide reaching applications in cosmology \cite{Ntampaka:2019udw} and the physical sciences more broadly \cite{Carleo:2019ptp}, now having been applied to problems in large-scale structure \cite{ Pan:2019vky}, the CMB \cite{Mishra:2019mbm,Farsian:2020adf, Caldeira:2018ojb}, 21~cm \cite{Hassan:2019cal} and lensing studies both weak \cite{ 2010ApJ...720..639G, Nurbaeva:2014fsa,Schmelzle:2017vwd,Fluri:2019qtp} and strong \cite{ Hezaveh:2017sht,PerreaultLevasseur:2017ltk,Morningstar:2018ase,Morningstar:2019szx,canameras2020holismokes}. Recently, promising results have been achieved with supervised machine learning algorithms for identifying dark matter substructure properties with simulated galaxy-galaxy strong lensing images. These include applications of convolutional neural networks (CNNs) for inference of population level properties of substructure \cite{Brehmer:2019jyt}, classification of halos with and without substructure \cite{Alexander:2019puy, DiazRivero:2019hxf} and between dark matter models with disparate substructure morphology \cite{Alexander:2019puy}, as well as classifying between lenses with different subhalo mass cut-offs \cite{Varma:2020kbq}. In a similar spirit, \cite{2020arXiv200811577V} used a convolutional neural network to classify simulated astrometic signatures of a population of quasars as being consistent with the presence of dark matter substructure in the Milky Way.

A complimentary direction is the application of unsupervised machine learning techniques to the challenge of identifying dark matter substructure. This scenario differs fundamentally from supervised approaches in that there is no longer a need for a labeled training set. It is also fundamentally different from a Bayesian approach as one does not assume a model {\it a priori}. Unsupervised machine learning algorithms are designed in such a way that the underlying structure of the data set used for training is learned through clustering, generating new data, removing noise from data, and detection of anomalous data. Unsupervised machine learning techniques have been used in cosmology including applications of generative adversarial networks \cite{ Mishra:2019sep, Sadr:2020rje, Yoshiura:2020ioc, List:2020qdz}, variational autoencoders \cite{ Yi:2020xgq}, and support vector machines \cite{ Xu:2013bka, Hajian:2014pra, Jennings:2018eko}, among others.

In this work we present a new application of unsupervised machine learning in the context of identifying dark matter substructure with strong gravitational lensing. We apply various methods, including autoencoders, a type of unsupervised machine learning algorithm which has seen great success in other areas of physics \cite{ Hajer:2018kqm, DeSimone, Farina:2018fyg, Cerri:2018anq, DAgnolo:2018cun, Blance, Collins:2018epr, Khosa:2020qrz,Romao:2020ocr}, with additional applications in cosmology \cite{ Hoyle:2015rha}, to the challenge of anomaly detection (AD). We show that trained adversarial, variational, and deep convolutional autoencoders can identify data {\it with} substructure as anomalous for further analysis. Such anomalies can be followed-up with a Bayesian likelihood analysis or a dedicated supervised approach to identify the type of substructure observed and estimate its properties. We additionally show that our best performing anomaly detector is able to reach its full potential by comparing its performance to that of an ``optimal'' anomaly detector. We additionally show that unsupervised machine learning algorithms can serve as a useful first step in an analysis pipeline of strong lensing systems, in combination with supervised models.

The structure of this paper is as follows: In Sec.~II we present a short review of the theory of strong gravitational lensing; in Sec.~III we discuss dark matter substructure and expected signatures. In Sec.~IV we discuss the simulation of strong lensing images for training and in Sec.~V describe the models and their training. In Sec.~VI we present the unsupervised analysis of our data set, and in Sec.~VII the combined unsupervised and supervised results. We conclude with a discussion and future steps in Sec.~VIII.

\section{Strong Lensing Theory}

\begin{figure}
    \centering
    \includegraphics[width=\linewidth]{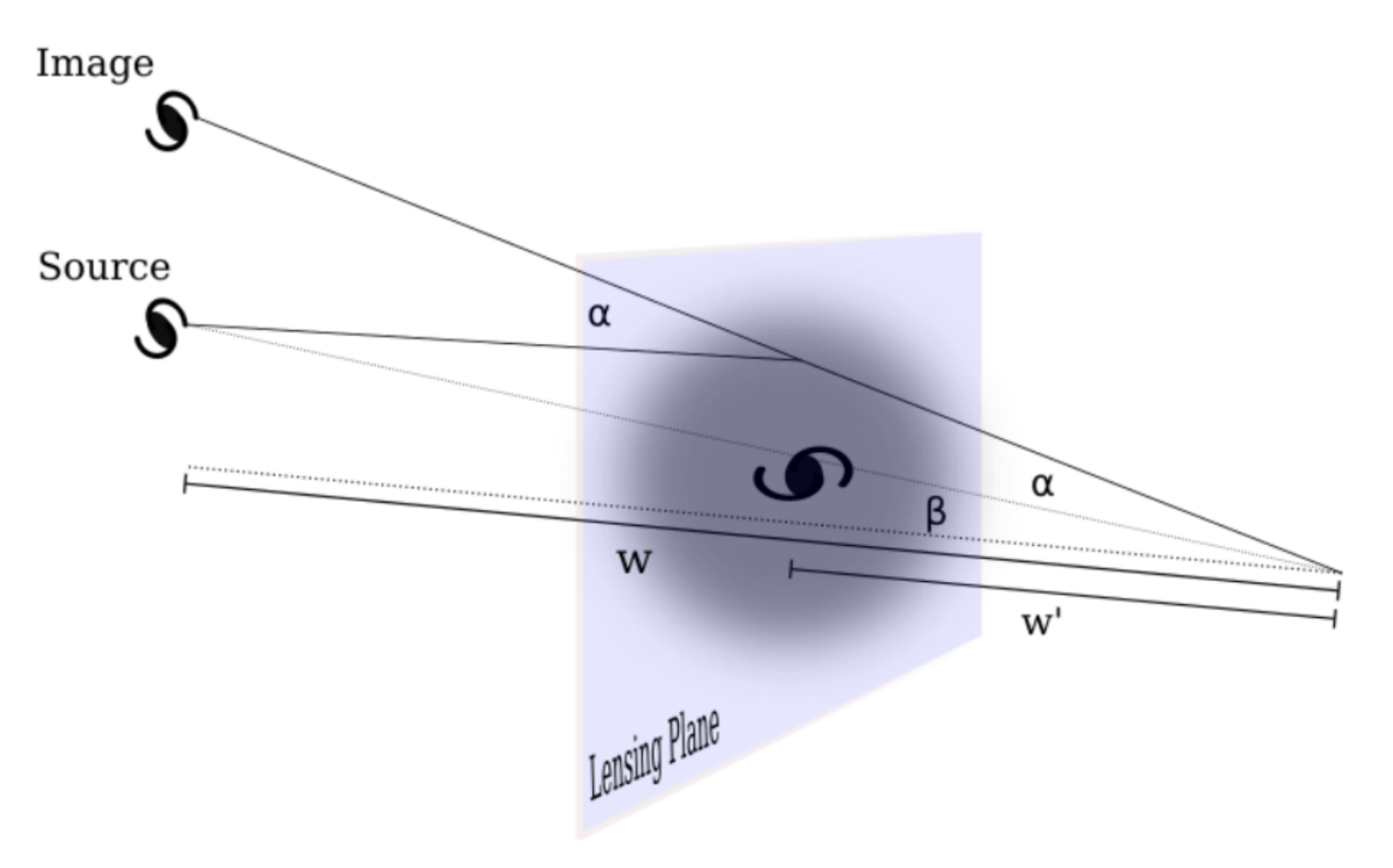}
    \caption{Schematic depicting lensing of a background galaxy by an intermediate dark matter halo per the {\it thin lens approximation}.}
    \label{fig:lens-dia}
\end{figure}

In this section we briefly review some of the relevant theory of strong lensing. To calculate the effects of lensing, we will consider scalar perturbations induced by a dark matter halo on a packet of null geodesics traveling in a cosmological background (see \cite{1994CQGra..11.2345S} for details beyond what is presented here). The evolution of these neighboring null geodesics is governed by the equation of geodesic congruence,
\begin{equation}
    \partial^k\partial_k  S^i = R S^i,
    \label{egc}
\end{equation}
where $S^i = \frac{dx^i}{ds}$ are tangent vectors for $i = 1, 2$ which are parallel transported along a fiducial ray $T^\mu = \frac{dx^\mu}{dt}$ such that $\nabla_k S^i = 0$ and $R$ is the optical tidal matrix. It is easy to show that the equation of geodesic congruence for a Friedmann--Lemaitre--Robertson--Walker (FLRW) geometry
reduces to,
\begin{equation}
    \left(\frac{d^2}{dw^2} + K\right)S^i = 0,
\end{equation}
where $K$ is the spatial curvature of the FLRW spacetime and $w$ is the comoving distance.

The effect of a dark matter halo in a FLRW spacetime can be implemented as a perturbation. The perturbed metric in Newtonian gauge is given by,
\begin{equation}
    ds^2 = -(1 + 2\phi)d\eta^2 + (1 - 2\phi)dw^2.
\end{equation}
We can now calculate the optical tidal matrix and use Equation \ref{egc} to calculate the perturbed equation of geodesic congruence. Working in the first order formalism, the obvious choice (under the assumption of a small lensing potential) of basis one forms is,
\begin{equation}
\begin{split}
        e^0 = (1 + \phi)~d\eta,\\
        e^i = (1 - \phi)~dw^i.
\end{split}
\end{equation}
The spin connection $w_{ab}$ can be constructed such that the first structure equation is satisfied,\footnote{Note that we use the convention where $,$ is shorthand for a derivative - i.e. $\phi_{,i} = \partial_i \phi$.}
\begin{equation}
\begin{split}
        w^0_i = \phi_{,i} ~ e^0,\\
        w^i_j = \phi_{,i} ~ e^j - \phi_{,j} ~ e^i,
\end{split}
\end{equation}
and the second structure equation yields the curvature 2--form $\Omega^a_b$,
\begin{equation}
\begin{split}
    \Omega^0_i = \phi_{,ik}e^k \wedge e^0, \\
    \Omega^i_j = -\phi_{,jk}e^k \wedge e^i + \phi_{,ik}e^k \wedge e^j.
\end{split}
\end{equation}
It is now trivial that the curvature scalar is given by $R = 2\partial^i \partial_j \phi$. Thus our perturbed equation of geodesic congruence is given by,
\begin{equation}
    \left( \frac{d^2}{dw^2} + K\right)x^i = -2\phi_{,i}.
    \label{pert-egc}
\end{equation}

Equation \ref{pert-egc} has a non--trivial solution for the comoving transverse distance, $x^i$. The Green's function can be calculated for initial values $x^i = 0$, $dS^i/dw = e^i \text{ at } w = 0$ and has a well established solution,
\begin{equation}
    x^i = f_K(w)e^i - 2\int_0^w dw' f_K(w - w') \phi_{,i}~,
\end{equation}
where $f_K$ is the angular--diameter distance in FLRW spacetime. Rearranging terms we arive at the well known {\it lens equation},
\begin{equation}
    \beta^i = e^i - 2\int_0^w dw' \frac{f_K(w - w') \phi_{,i}}{f_K(w)f_K(w')}.
    \label{eqlens}
\end{equation}
The second term on the r.h.s side of Equation \ref{eqlens} is known as the deflection angle $\alpha^i$ which is related to the effective lensing potential $\psi$ by the angular, or perpendicular, gradient $\psi_{;\perp} = \alpha$.

$\kappa$, the convergence, or measure of the mass density in the lensing plane, is a useful parameter -- see Figure \ref{fig:lens-dia} for a schematic for strong lensing. Under this {\it thin lens approximation} the effective lensing potential can be written as,
\begin{equation}
    \psi = \frac{1}{\pi}\int d^2\theta'\kappa(\theta')\ln{\left| \theta - \theta'\right|},
\end{equation}
where $\theta = e^i$ is the position on the lensing plane. Interestingly, in this form we can find a Poisson equation for gravitational lensing where the the convergence can be thought of as if it where charge in electromagnetism. Taking the two-dimensional Laplacian and given that we are free to move the differentiation inside the integral,
\begin{equation}
    \nabla^2\psi =  \frac{1}{\pi}\int d^2\theta'\kappa(\theta')\nabla^2\ln{\left| \theta - \theta'\right|}.
    \label{lap-kap}
\end{equation}
The logarithm in Equation \ref{lap-kap} is simply the Green's function for a two--dimensional Laplacian, thus we can write,
\begin{equation}
    \nabla^2\psi =  2 \int d^2\theta'\kappa(\theta')\delta(\theta - \theta'),
\end{equation}
which simplifies to,
\begin{equation}
    \nabla^2\psi = 2\kappa,
    \label{poiss}
\end{equation}
which is the Poisson equation for gravitational lensing. 

\section{Dark Matter Substructure}

The $\Lambda$ Cold Dark Matter ($\Lambda$CDM) model predicts that nearly scale invariant density fluctuations present in the early universe evolve to form large scale structure through hierarchical structure formation. This model envisions dark matter halo formation originating from the coalescence of smaller halos \cite{Kauffmann:1993gv} with N-body simulations predicting that evidence of this merger history be observable given that subhalos should avoid tidal disruption and remain largely intact. The distribution of subhalo masses can be well modeled with power law distribution,
\begin{equation}
    \frac{dN}{d m} \propto m^{\beta},
    \label{subdist}
\end{equation}
where $\beta$ is known to be $\sim - 1.9$ from simulations \cite{Springel:2008cc, Madau:2008fr}. While $\Lambda$CDM has seen great success on large scales being consistent with the cosmic microwave background, galaxy clustering, and weak lensing  \cite{planck,clust,wlens}, subgalactic structure predictions inferred by $\Lambda$CDM have come under scrutiny. Some well known issues with this model include the missing satellites \cite{Moore_1999,Klypin_1999,msp} (the number of observed subhalos doesn't align with observation)\footnote{See \cite{mspr} for a differing perspective.}, core-vs-cusp (rotations curves are found to be cored \cite{burk,Oh_2015} and not cuspy as expected from simulation \cite{nfw}), diversity (the profile of inner regions of galaxies is more diverse than expected \cite{Oh_2015}), and too big to fail problems (brightest subhalos of our galaxy have lower central densities than expected from N-body simulations \cite{Boylan_Kolchin_2011}). In light of this, it is prudent to consider other promising models of dark matter.

In addition to the well studied case of substructure from non-interacting particle dark matter, one can also extend to other well motivated theories like dark matter condensates which constitute both Bose-Einstein (BEC) \cite{Sin:1992bg,Silverman:2002qx,Hu:2000ke,Sikivie:2009qn,Hui:2016ltb,Berezhiani:2015bqa,Ferreira:2018wup} and Bardeen-Cooper-Schreifer (BCS) \cite{Alexander:2016glq,Alexander:2018fjp}.  A leading example of condensate dark matter is axion dark matter. Axions were first introduced as a solution to the strong-CP problem of the Standard Model \cite{Peccei:1977hh,Wilczek:1977pj,Weinberg:1977ma} and were later proposed as a possible form of dark matter\cite{Preskill:1982cy,Abbott:1982af,Dine:1982ah}. Furthermore, as a Goldstone boson of a spontaneously broken U(1) symmetry, they are by construction the field theory definition of superfluidity \cite{Schmitt:2014eka}. Interestingly, for a specific choice of effective field theory, one can reproduce the baryonic Tully-Fisher relation \cite{Berezhiani:2015bqa,Berezhiani:2015pia}.

This type of dark matter can form quite exotic substructure like vortices \cite{sfdm} and disks \cite{Alexander:2019qsh}. The vortex density profile was studied by \cite{sfdm} and parameterized as a tube,
\begin{equation}
    \rho_{v}(r,z)= 
\begin{cases}
    0 ,&  r > r_{v}\\
    \rho_{v0}\left[ \left(\frac{r}{r_{v}}\right)^{\alpha_v}-1 \right],              &  r \leq r_v
\end{cases},
\end{equation}
where $r$ is the radial distance, $r_v$ is the core radius, and $\alpha_v$ is a scaling exponent. On large scales, one can simplify the vortex model as a linear mass of density $\rho_{v0}$. The values for these parameters, as well as the expected number density in realistic dark matter halos, varies across the literature. The number of expected vortices in halos range from 340 in the M31 halo for particle mass $m = 10^{-23}$ eV \cite{Silverman:2002qx} to $N = 10^{23}$ for a typical dark matter halo with $m = 1$ eV \cite{Berezhiani:2015bqa}. Interestingly, it was shown in \cite{Banik:2013rxa} that vortices have a mutual attraction and could combine to form a single, more massive vortex over time. 

While the lensing for subhalos is well understood, lensing effects from more exotic substructure like superfluid vortices has not been explicitly studied. However, vortices can be thought of as a non-relativistic analog of cosmic strings which form during a phase transition in a quantum field theory \cite{Brandenberger:1993by,Brandenberger:2013tr}. Studies of lensing from cosmic strings in the literature \cite{Sazhin:2006kf,Gasparini:2007jj,Morganson:2009yk} are carried out under the simplifying assumption that the cosmic string velocity is non-relativistic which corresponds precisely to studying lensing from a vortex. Qualitatively, the lensing from vortices is similar to subhalos in that they can produce multiple images but differ in that there is no magnification of the background source. 

A final note is the effect that line of sight halos, also known as interlopers, might have on the lensing signature. It is expected that interlopers should have a non-negligible contribution to the distortion of gravitational lenses and may even dominate the signature due to substructure \cite{Sengul:2020yya,McCully_2017,Despali_2018,Gilman_2019}. Thus, careful studies of substructure in dark matter halos should account for their influence.

\section{Strong Lensing Simulations} \label{sec:SLS}

\begin{table}
    \scriptsize
	\centering
	\caption{Parameters with distributions and priors used in the simulation of strong lensing images. Note that only a single type of substructure was used per image.}
	\label{tab:table}
	\begin{tabular}{cccc}
		\hline
		\hline
		DM Halo\\ 
		\hline
		\hline
		\textbf{Param.} & \textbf{Dist.} & \textbf{Priors} & \textbf{Details} \\
		\hline
$\theta_x$  &  fixed & 0 & x position \\
$\theta_y$  &  fixed & 0 & y position \\
z  & fixed & 0.5  & redshift\\
$M_{\rm TOT}$ & fixed & 1e12 & total halo mass in $ {\rm M}_\odot$\\
        \hline
		\hline
		Ext. Shear\\
		\hline
		\hline
		\textbf{Param.} & \textbf{Dist.} & \textbf{Priors} & \textbf{Details} \\
		\hline
		$\gamma_{ext}$ & uniform & [0.0, 0.3] & magnitude \\
		$\phi_{ext}$ & uniform & [0, 2$\pi$] & angle \\
        \hline
		\hline
		Lensed Gal.\\ 
		\hline
		\hline
		\textbf{Param.} & \textbf{Dist.} & \textbf{Priors} & \textbf{Details} \\
		\hline
$r$ & uniform & [0, 0.5] & radial distance from center\\
$\phi_{bk}$ & uniform & [0, 2$\pi$] & orientation from y axis\\ 
z  &  fixed & 1.0  & redshift\\
e  &  uniform & [0.4, 1.0] & axis ratio\\
$\phi$ & uniform & [0, 2$\pi$] & orientation to y axis \\

n & fixed & 1.5 & Sersic index \\
R & uniform & [0.25,1] & effective radius \\
        \hline
		\hline
		Vortex \\
		\hline
		\hline
		\textbf{Param.} & \textbf{Dist.} & \textbf{Priors} & \textbf{Details} \\
		\hline
$\theta_x$  &  normal & $[0.0,0.5]$ & x position \\
$\theta_y$  & normal & $[0.0,0.5]$ & y position\\
$l$ & uniform & [0.5,2.0] & length of vortex\\
$\phi_{vort}$ & uniform & [0, 2$\pi$] & orientation from y axis\\ 
$m_{vort}$ & uniform & [3.0,5.5] & \% of mass in vortex\\
        \hline
		\hline
		Subhalo \\
		\hline
		\hline
		\textbf{Param.} & \textbf{Dist.} & \textbf{Priors} & \textbf{Details} \\
		\hline
$r$ & uniform & [0, 2.0] & radial distance from center\\
$N$ & Poisson & $\mu$=25 & number of subhalos\\
$\phi_{sh}$ & uniform & [0, 2$\pi$] & orientation from y axis\\ 
$m_{sh}$ & power law & [1e6,1e10] & subhalo mass in ${\rm M}_{\odot}$\\
$\beta_{sh}$ & fixed & -1.9 & power law index

	\end{tabular}
\end{table}

There is currently only a small number of observed and identified strong galaxy-galaxy lensing images. Nonetheless, it is very timely to study and benchmark different algorithms with simulated strong lensing images, before the influx of data from Euclid and the Vera C. Rubin Observatory, where we can expect thousands of high quality lensing images \cite{2019arXiv190205141V,2010MNRAS.405.2579O}. 
\begin{figure}
    \centering
    \includegraphics[width=0.7\linewidth]{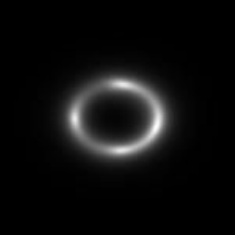}
    \caption{A sample strong lensing image simulated with PyAutoLens.}
    \label{fig:lens_examp}
\end{figure}

For our analysis we simulate strong lensing data without substructure and substructure from two disparate types of dark matter - subhalos of CDM and vortices of superfluid dark matter using the \texttt{PyAutoLens} package \cite{2015MNRAS.452.2940N,2018MNRAS.478.4738N}. The parameters used in the simulation are shown in Table \ref{tab:table}. Images are composed $150 \times 150$ pixels with a pixel scale of $0.5''/$pixel (as shown in an example image in Figure \ref{fig:lens_examp}). Informed by real strong galaxy-galaxy lensing images, we further include background and noise in our simulations such that the lensing arcs have a maximum signal-to-noise ratio (SNR) of $\sim 20$ \cite{hst}. We further include modifications induced by a point-spread function (PSF) approximated by an Airy disk with a first zero-crossing at $\sigma_{\rm psf} \lesssim 1''$. We model the light from lensed galaxies with a basic Sersic profile. 

In simulating substructure, we approximate subhalos as point masses, drawing their masses from Equation \ref{subdist}. We determine the number of subhalos for each image from a Poisson draw for a mean of 25, which is consistent with the number of expected subhalos for our field of view at our range of redshifts \cite{Rivero:2018bcd}. Vortices are approximated as uniform density strings of mass of varying length. During simulation we ensure that the total mass of the main dark matter halo plus the mass of substructure is always equal to $1 \times 10^{12} ~ {\rm M}_\odot$. This is to ensure that the models don't simply recognize that simulations without substructure are less massive. Our simulations correspond to a total fraction of mass in substructure of $\mathcal{O}(1\%)$. In addition to the effects of substructure of the dark matter halo, we further include the effects induced by external shear due to large-scale structure.
The linearity of the Poisson equation, Equation \ref{poiss}, implies the total lensing is the sum of the separate contributions. Explicitly,
\begin{equation}
    \alpha = \alpha_{LSS}  + \alpha_{halo} + \alpha_{halo-sub},
\end{equation}
where $\alpha_{LSS}, ~\alpha_{halo}, ~\alpha_{halo-sub}$ are the external shear from large-scale-structure and lensing from the halo and halo substructure. Thus the location of an image calculated in our simulations is given by the following modified form of the \textit{lens equation}, Equation \ref{eqlens},
\begin{equation}
    \beta^i = \theta^i - \alpha^i_{LSS}  - \alpha^i_{halo} - \alpha^i_{halo-sub}.
\end{equation}

\section{Network Architecture \& Training}

\subsection{Network Architectures}
For the supervised approach, we consider \textit{ResNet-18} \cite{2015arXiv151203385H} and \textit{AlexNet} \cite{Krizhevsky:2014} as baseline architectures. These are the same architectures used in \cite{Alexander:2019puy} to perform multi-class classification of simulated strong lensing images with differing substructure.

For the unsupervised approach, we consider three types of autoencoder models \cite{Rumelhart:1986} and a variant of a Boltzmann machine \cite{Hinton:1983}. The goal of an autoencoder neural network is to learn self-representation of the input data. They consist of an encoder network and a decoder network. The encoder learns to map the input samples to a latent vector whose dimension is lower than that of the input samples, and the decoder network learns to reconstruct the input from the latent dimension. The restricted Boltzmann machine (RBM) is realized as a bipartite graph that learns a probability distributions for inputs \cite{2018arXiv180607066M}. RBMs consist of two layers, a hidden layer and a visible layer, where training is done in a process called contrastive divergence \cite{hint}. 

We first consider a deep convolutional autoencoder \cite{10.1007/978-3-642-21735-7_7}, which is primarily used for feature extraction and reconstruction of images. During training we use the mean squared error (MSE),
\begin{equation}
    {\rm MSE}(\hat{\theta}) = \mathbb{E}_\theta\left[ \left( \theta - \hat{\theta} \right)^2 \right],
\end{equation}
as our reconstruction loss where $\theta$ and $\theta'$ are the real and reconstructed samples. See Table \ref{tab:table-DCAE} in the Appendix for details of the deep convolutional autoencoder model.

We next consider a variational autoencoder \cite{2013arXiv1312.6114K} which introduces an additional constraint on the representation of the latent dimension in the form of Kullback-Liebler (KL) divergence,
\begin{equation}
    D_{KL}\left( P || Q \right) = \mathbb{E}_{x \sim P}\left[ - \log{ {Q(x)} }\right] - \mathcal{H}(P(x)),
\end{equation}
where $P(x)$ is the target distribution and $Q(x)$ is the distribution learned by the algorithm. The first term is the cross entropy between $P$ and $Q$ and the second term is the entropy of $P$. In the context of variational autoencoders, the KL divergence serves as a regularization to impose a prior on the latent space. $P$ is chosen to take the form of a Gaussian prior on the latent space $z$ and $Q$ corresponds to the approximate posterior $q(z|x)$ represented by the encoder. The total loss of the model is the sum of reconstruction (MSE) loss and the KL divergence. We implement KL cost annealing where only the reconstruction loss is used during the first 100 epochs and then the weight of KL divergence loss is gradually increased from 0 to 1. The details of the variational autoencoder architecture are presented in Table \ref{tab:table-VAE} of the Appendix.

\begin{figure}[!t]
    \centering
    \includegraphics[width=\linewidth]{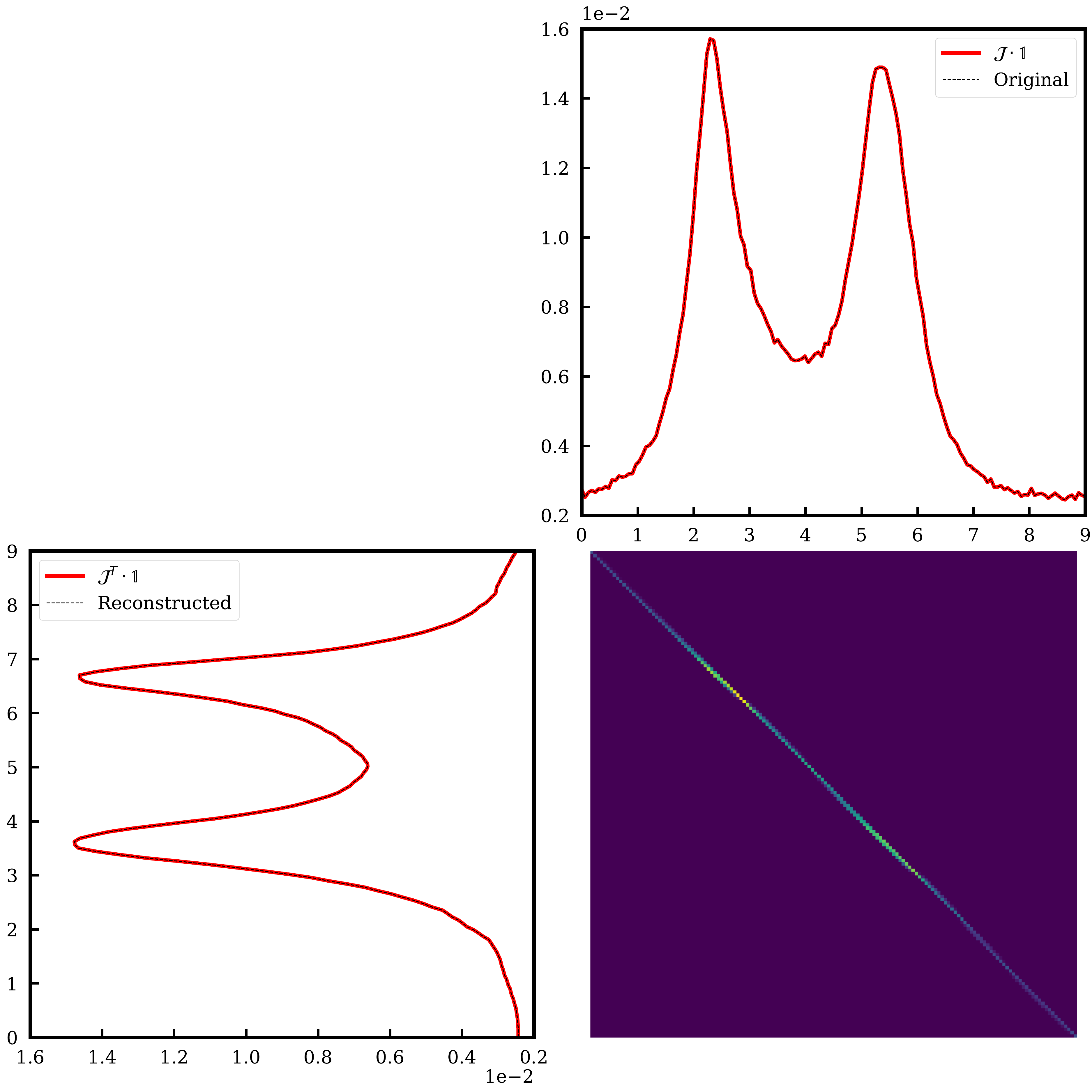}
    \caption{Optimal transport matrix between the real image (top right) and the reconstructed image (bottom left).}
    \label{fig:wass-examp}
\end{figure}

We additionally consider an adversarial autoencoder \cite{2015arXiv151105644M} which replaces the KL divergence of the variational autoencoder with adversarial learning. We train a discriminator network $D$ to classify between the samples generated by the autoencoder $G$ and samples taken from a prior distribution $P(z)$ corresponding to our training data. The total loss of the model is the sum of reconstruction (MSE) loss and the loss of the discriminator network,
\begin{equation}
    L_D = \mathbb{E}_{x \sim p_{\rm data}}\left[ \log(D(x)) \right] + \mathbb{E}_{z \sim P}\left[ \log(1 - D(G(z))) \right].
\end{equation}
A regularization term is added to the autoencoder of the following form,
\begin{equation}
    L_G =  \mathbb{E}_{z}\left[ \log(D(z)) \right].
    \label{genloss-eq}
\end{equation}
As the autoencoder becomes proficient in reconstruction of inputs the ability of the discriminator is degraded. The discriminator network then iterates by improving its performance at distinguishing the real and generated data. Details of the adversarial model architecture are presented in Table \ref{tab:table-AAE} of the Appendix.

\subsection{Network Training and Performance Metrics}

For training supervised architectures we use $25,000$ training and $2,500$ validation images per class. The cross-entropy loss was minimized with the {\it Adam} optimizer in batches of 250 for 50 epochs. The learning rate was initialized at $1 \times 10^{-3}$ and allowed to decay at an increment of $1 \times 10^{-5}$ every epoch. We implement our architectures with the \texttt{PyTorch} \cite{NEURIPS2019_9015} package run on a single NVIDIA Tesla P100 GPU. Similarly, we use 25,000 samples with no substructure and 2,500 validation samples per class for training the unsupervised models. The models are implemented using the \texttt{PyTorch} package and are run on a single NVIDIA Tesla K80 GPU for 500 epochs. 

For training the AAE, the autoencoder and discriminator are trained alternatively (their parameters are not updated at the same time) but in the same iteration steps. The encoder's output is used as an input to the discriminator, and there is no relative importance or weighting explicitly given to the discriminator loss or MSE loss, and all gradients are calculated independently. The MSE loss is calculated first for updating both the encoder and the decoder parameters, then the discriminator loss is calculated for updating the discriminator parameters. Lastly, the entropy loss is calculated for updating the encoder parameters, i.e. regularization.

We utilize the area under the ROC curve (AUC) as a metric for classifier performance. For unsupervised models, the ROC values are calculated for a set threshold of the reconstruction loss.
In calculating the ROC, true and false positives are determined based on the \textit{known} label of the input. For example, an image with no substructure (substructure) will be counted as a false negative (true negative) if its loss is above the threshold.
We additionally use the Wasserstein distance metric described below as an additional cross-check performance metric for unsupervised models. 

The {\it Wasserstein distance}, a metric defining the notion of ``distance'' between probability distributions, is briefly described below. For a metric space $M$ there exists $X \sim P$ and $Y \sim Q$, for $X, Y \in \mathbb{R}^d$, with probability densities $p$, $q$. The p$^{\rm th}$-Wasserstein distance is given by,
\begin{equation}
    W_p(P,Q) := \left( \mathop{{\rm inf}}_{J \in \mathcal{J}(P,Q)} \int || x - y ||^p dJ(x,y) \right)^{\frac{1}{p}},
    \label{Wp}
\end{equation}
for $p \geq 1$, where the infimum is taken over the space of all joint measures $J$ on $M \times M$ denoted by $\mathcal{J}(P,Q)$. The special case of $p = 1$ corresponds to the well known {\it Earth Mover distance}. We can write Equation~\ref{Wp} down in a form that is useful for empirical data,
\begin{equation}
     W_p(P,Q) := \mathop{{\rm min}}_{J} \left( \sum_{i,j} ||x_i - y_j ||^p J_{ij} \right)^\frac{1}{p},
     \label{wass-eq}
\end{equation}
where $J_{ij}$ is known as the optimal transport matrix and has the property that,
\begin{equation}
    J \cdot \mathbb{1} = p {\rm   ~~~ \& ~~~ } J^{T} \cdot \mathbb{1} = q
\end{equation}

We use the Wasserstein distance to measure the deviation between the input and reconstructed images for unsupervised algorithms. In practice, this is achieved by compressing images down an axis to a 1D representation and then solving Eq.~\ref{wass-eq} for the optimal transport matrix $J_{ij}$ for the geodesic between both images, as shown in Figure \ref{fig:wass-examp}. Thus, the architecture with the shortest average Wasserstein distance can be interpreted as the best reproduction of a given data set. We calculate the Wasserstein distance based on the average compression down the $x$ and $y$ axis.

\begin{figure}
    \centering
    \includegraphics[width=\linewidth]{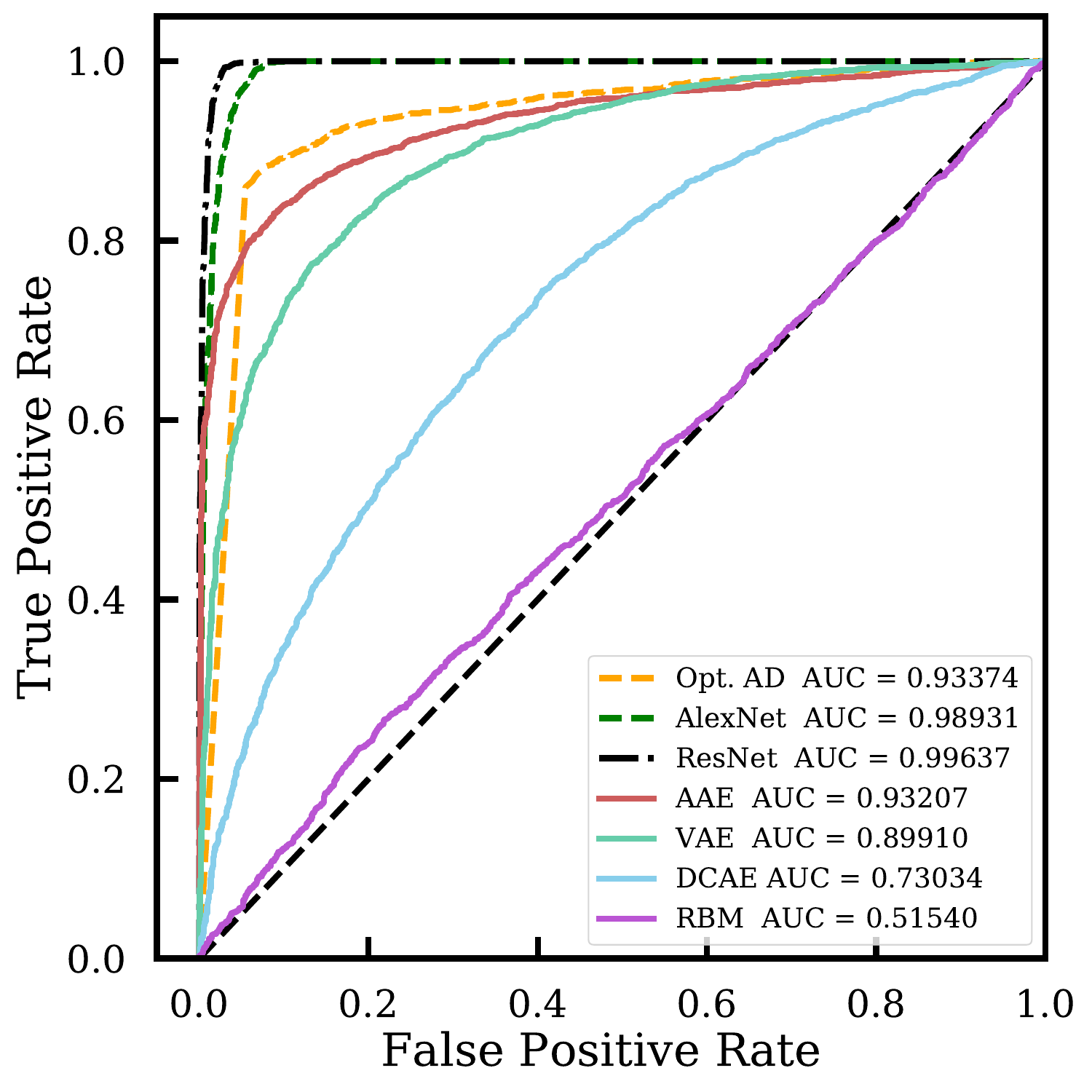}
    \caption{ROC-AUC curve for unsupervised and supervised algorithms applied to our data set.}
    \label{fig:unsup-roc-auc}
\end{figure}

\begin{table}[]
    \centering
    \begin{tabular}{l c c}
	\hline
	\hline
	\textbf{Architecture} \hspace{.7cm}& \textbf{AUC} & \hspace{1cm} \textbf{W$_{1}$}\\
    \hline
    {ResNet-18} \hspace{.7cm} &  0.99637 &  \\
    {AlexNet} \hspace{.7cm} &  0.98931 &  \\
     AAE \hspace{.7cm}            &  0.93207 & \hspace{1cm} 0.22112\\
     VAE  \hspace{.7cm}          &  0.89910  & \hspace{1cm} 0.22533\\
     DCAE \hspace{.7cm}            &  0.73034 & \hspace{1cm} 0.26566\\
     RBM     \hspace{.7cm}        &  0.51054 & \hspace{1cm} 1.27070\\
    {ResNet-18 (AD)} \hspace{.7cm} &  0.93374 &  \\
    \end{tabular}
    \caption{Performance of architectures used in this analysis. AUC values for supervised architectures are calculated for classification between images with and without substructure - thus it is not a macro-averaged AUC. ResNet-18 (AD) corresponds to our \textit{ResNet} architecture trained as an ``optimal'' anomaly detector. $W_1$ is the average $1^{\rm st}$ Wasserstein distance for images without substructure. See Table \ref{tab:complete-wass} in the Appendix for the set of all Wasserstein distances.}
    \label{tab:arc_tab}
\end{table}

\begin{figure*}[!t]
    \centering
    \includegraphics[width=\linewidth]{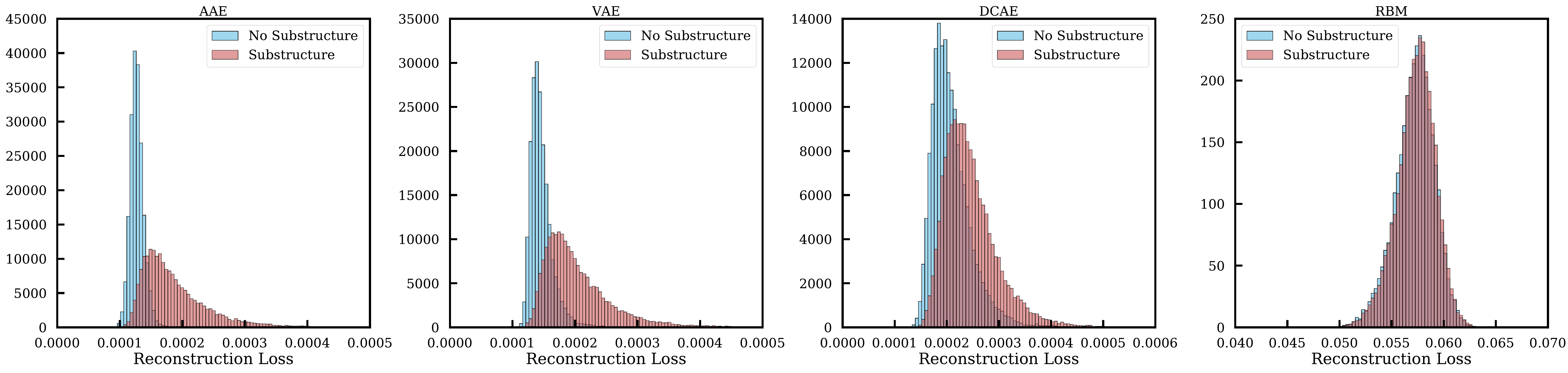}
    \caption{Reconstruction loss for  architectures. Left to right: AAE, VAE, DCAE, and RBM.}
    \label{recon-loss-A}
\end{figure*}

\section{Results}

Four different unsupervised architectures were studied in the context of anomaly detection - a deep convolutional autoencoder (DCAE), convolutional variational autoencoder (VAE), adversarial autoencoder (AAE) and a restricted Boltzmann machine (RBM). The results for all four architectures are shown in Figure \ref{fig:unsup-roc-auc}, together with the supervised results. Model performance metrics are collected in Table~\ref{tab:arc_tab} and complete architecture layouts are presented in the Appendix.

The RBM model has the poorest performance overall, which is expected as its architecture is not optimal for image inputs. Although the RBM model does not distinguish well between images with differing substructure, its low reconstruction loss implies that it does succeed in reconstructing the general morphology of the lens.
Three autoencoder based architectures do significantly better. The DCAE model achieves an AUC of $\approx 0.730$, the VAE model $\approx 0.899$, and the AAE model achieves top performance, nearing supervised discriminating power, at  $\approx 0.932$. These improvements are also clearly visible in the distribution of reconstruction losses in Figure \ref{recon-loss-A}.

We stress here that it is not expected that the fully unsupervised model should attain the performance of the supervised model on labeled data, which would be extremely difficult, given the advantage of the supervised model that is trained to distinguish the specific substructure models under study. However, despite only being trained on the null (non-substructure) class, the best unsupervised model (AAE) achieves remarkably high performance compared to the fully-supervised models.

To further study just how well the unsupervised architectures perform, we train a nearly optimal anomaly detector. We do this by constructing a classifier trained to distinguish no substructure from no substructure + substructure class using the same network parameters as the supervised model. This model achieves an AUC of $\approx$  0.934, implying that the AAE has near optimal performance as an anomaly detector. This demonstrates that the unsupervised model can perform as a robust general substructure anomaly detector with the benefit that it is not tied to any specific substructure model.

\begin{figure*}
    \centering
    \includegraphics[width=\linewidth]{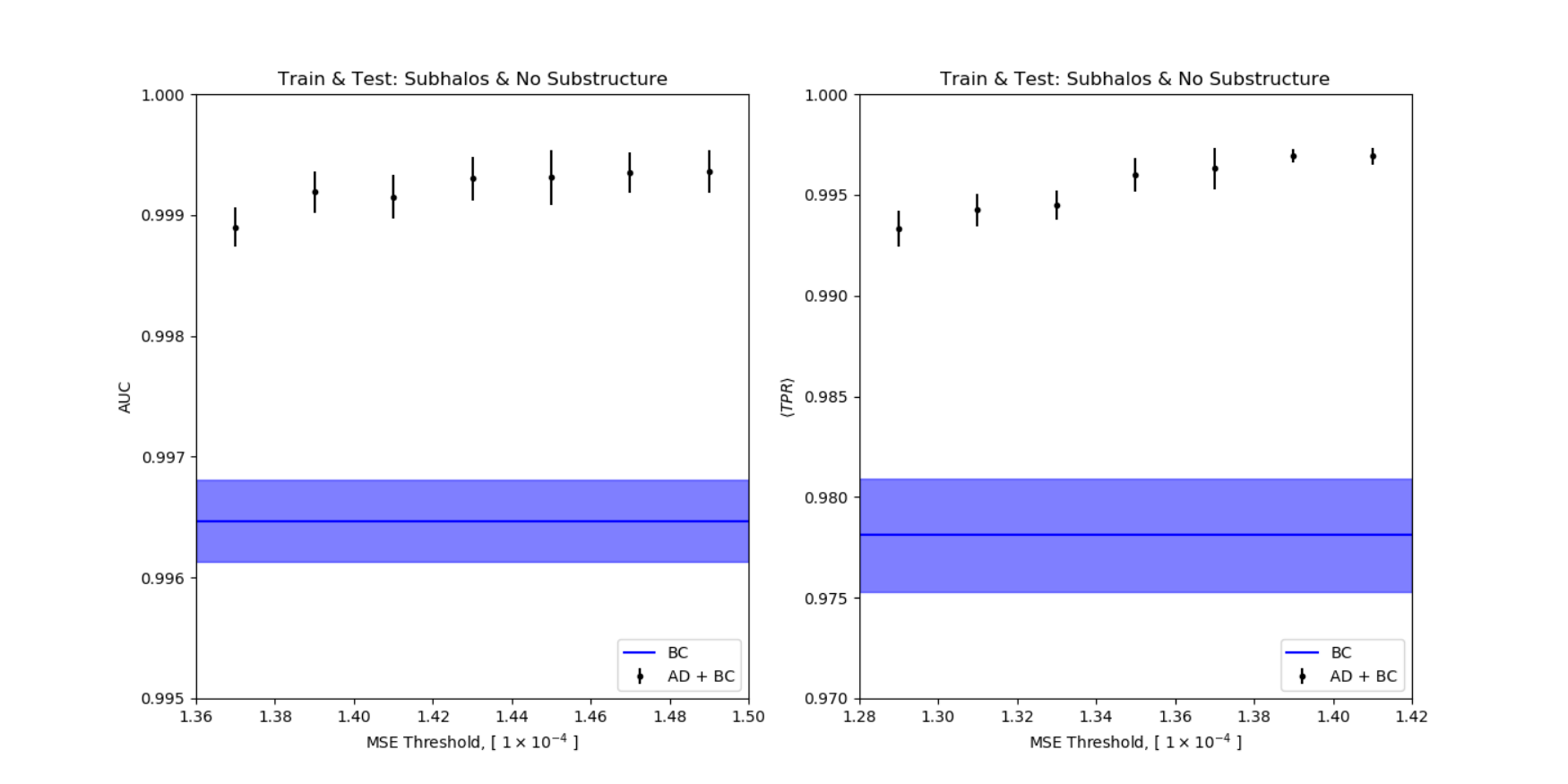}
    \caption{AUC Performance (left) and average TPR over FPR [0,0.2] (right), for binary classifier and combined BC/AD algorithms.}
    \label{fig:subhalo-com}
\end{figure*}

As an additional performance metric for unsupervised architectures we calculate the average Wasserstein distance from our validation data set. We first do this for the no substructure class as shown in Table \ref{tab:arc_tab}. As the Wasserstein distance is a geodesic between distributions, smaller values correspond to distributions that are more similar. The values compiled in Table \ref{tab:arc_tab} show that the AAE and VAE achieve the best performance and that of the DCAE is $\sim$ 20\% larger by comparison. The ability of the RBM to reconstruct the inputs is significantly degraded compared to the autoencoder models. Together, these results are consistent with calculated AUC values for the architectures.

It is interesting to also consider the Wasserstein distance for data with substructure - i.e. a class that the models were not trained on. These results are compiled in Table \ref{tab:complete-wass} of the Appendix. All the autoencoders consistently have smaller distances for no substructure compared to that calculated from substructure. Furthermore, the AAE and VAE seem to be slightly better at reconstructing images for subhalo substructure versus vortices. This may be a result of higher symmetry from the effects of subhalo substructure.

In Appendix A we also identify the distribution of reconstruction losses for vortices and subhalos separately in Figure \ref{fig:recon-loss-sub-a}. As expected, there doesn't appear to be significant constraining power between the two types of substructure. The unsupervised models used as anomaly detectors can accurately distinguish between the no substructure and substructure scenarios, something they are designed for, but do not accurately distinguish between the different types of substructures (ie. differing types of anomalies). This is a more natural task for dedicated supervised machine learning algorithms, or for a combination of unsupervised and supervised algorithms.

\begin{figure*}[!t]
    \centering
    \includegraphics[width=\linewidth]{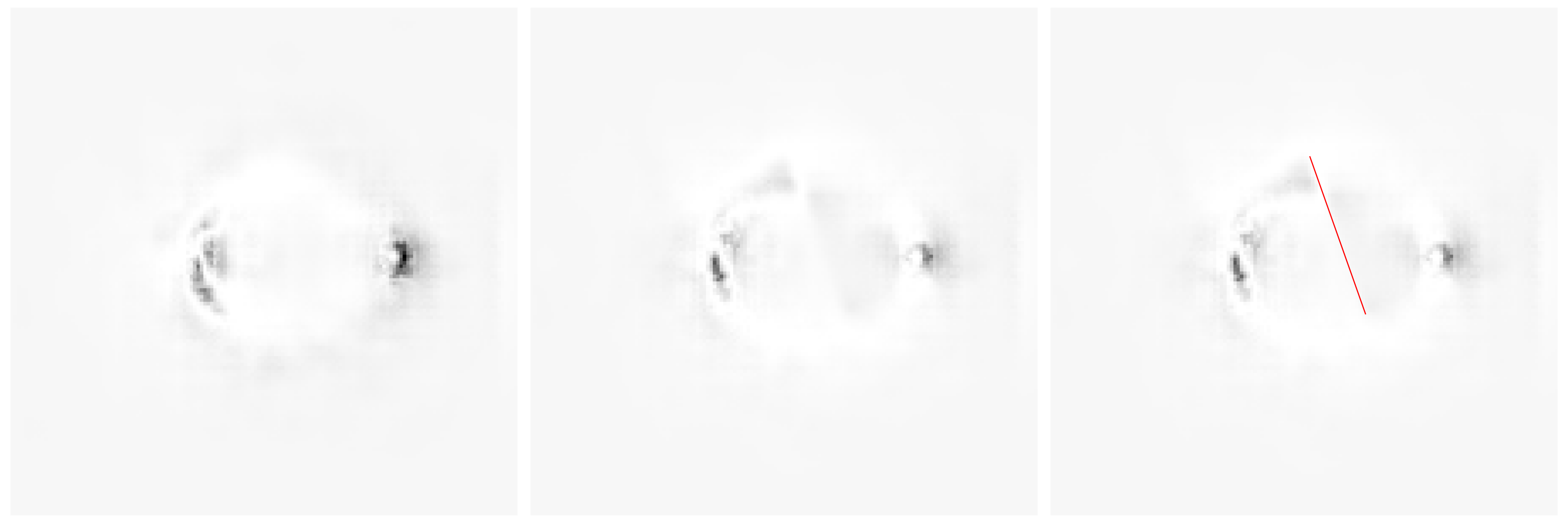}
    \caption{MSE loss for identical simulated images (save substructure) with no substructure (left), vortex substructure (middle), and vortex substructure with labeled vortex position (right).}
    \label{fig:vort_mse}
\end{figure*}

\section{Combination of Unsupervised and Supervised Models}

To test the benefits of combined unsupervised and supervised models, we test the performance of an analysis pipeline consisting of a sequence of unsupervised AAE and supervised {\it ResNet}. The AAE is used to isolate the most anomalous data which is then passed on to the supervised model for classification. Figure \ref{fig:subhalo-com} shows the results for identifying subhalos. The inclusion of the autoencoder as a first step appreciably improves the performance of the supervised classifier alone.\footnote{Note that when AUC values are so close to unity, even small increases in performance are significant. For a more challenging scenario, the improvements can become even more significant.}

We also investigated the reconstruction loss for images with and without substructure for the top performing unsupervised AAE model trained on the no substructure class. Interestingly, we find that the MSE loss for our data appears to encode information on the location of substructure. This effect is visualized in Figure \ref{fig:vort_mse} for a vortex of known location (see also Figure \ref{fig:sub-mse} in the Appendix for an example with subhalos). An algorithm that can identify deviations in the lensing image produced solely by its substructure is quite interesting. If this were possible, one could imagine using this data to invert the lens equation to produce the distribution of substructure mass on the lensing plane, which we leave to future work.

As a final cross-check, we investigate the possibility of classifying substructure based on unsupervised error maps directly. We train three {\it ResNet} architectures as binary  classifier between vortex, subhalo, and no substructure classes. The AUC between no substructure and subhalos was 0.95939, no substructure and vortices 0.99690, and between no substructure and both types of substructure 0.98361, comparable with binary classification of lensing images directly. The implies that the errors obtained through anomaly detection contain as much useful information about substructure as the images themselves.

\section{Discussion \& Conclusion}

It is now well known that dark matter substructure can be a useful probe to constrain models of dark matter based on their morphology. That can manifest observationally via their imprint on extended lensing arcs. It has been previously shown that machine learning has the power to identify dark matter models based on signatures unique to substructure \cite{Alexander:2019puy}. In this work we highlight the versatility of unsupervised machine learning algorithms in identifying the presence of dark matter substructure in lensing images, regardless of the substructure model.

We applied unsupervised models to realistic simulated data sets and found that a deep adversarial autoencoder achieves the best performance for detecting substructure anomalies, when compared to other types of autoencoders and a restricted Boltzmann machine, nearly saturating the optimal anomaly detector performance. By calculating the average Wasserstein distances for the data we further quantified the performance of these architectures.

The results support the conclusion that unsupervised models can be used as model-independent dark matter substructure anomaly detectors and a useful first step in an analysis pipeline to establish the most anomalous sources that can be further followed up with dedicated supervised algorithms or with standard Bayesian likelihood analysis techniques. We tested the performance of such a pipeline by combining an unsupervised and supervised model to classify data into substructure and no substructure classes for known substructure classes. The combined discriminating power of an autoencoder and a fully-supervised classifier was able to appreciably improve binary classification performance for known substructure classes. Finally, our anomaly detector is generally sensitive to other types of substructure models that are not currently known or anticipated, and hence can form an important standalone component of general dark matter substructure searches, while achieving performance comparable to supervised architectures.

We also found that calculating the reconstruction loss for images with substructure hints at their location, a useful property for future work. We found that the performance of binary classifiers trained on error map data was comparable to using lensing images directly, meaning that it contains all the relevant information for classification. While we do not explore error maps further in this work, they could prove useful in trying to \textit{invert} the lensing equation to infer the distribution of dark matter substructure on the lensing plane.

To our knowledge, this is the first result that shows that a fully unsupervised model can reach close to ideal anomaly detection performance in a realistic application, is competitive with supervised models in scenarios that favor supervised models by design, and when combined with supervised models gives the top performance overall. To this end, we believe that unsupervised algorithms will play a critical role in future machine learning applications to dark matter searches.

\section{Acknowledgements}
We would like to thank Evan McDonough and Ali Hariri for useful discussions. K Pranath Reddy is a participant in the Google Summer of Code (GSoC) 2020 program. S. G. was supported in part by the National Science Foundation Award No. 2108645. S. A. was supported in part by U.S. National Science Foundation Award No. 2108866. This work made use of these additional software packages: \textit{Matplotlib} \cite{Hunter:2007}, \textit{NumPy} \cite{harris2020array}, \textit{PyTorch} \cite{NEURIPS2019_9015}, and \textit{SciPy} \cite{2020SciPy-NMeth}.


%

\onecolumngrid

\pagebreak
\appendix
\section{Extra Tables \& Figures}

\begin{table}[h!]
    Wasserstein Distances\\ \vspace{0.1cm}
    \centering
    \begin{tabular}{l l c c c}
	\hline
	\hline
	&\textbf{AAE} \hspace{1cm}& \textbf{VAE} \hspace{1cm}& \textbf{DCAE}&  \hspace{1cm} \textbf{RBM}\\
    \hline
     no substructure\hspace{1cm} & 0.22403 \hspace{1cm} & 0.23029 \hspace{1cm} &  0.27125 & \hspace{1cm} 1.32840 \\
     subhalo \hspace{2cm} & 0.25761 \hspace{1cm} & 0.24948 \hspace{1cm}            &  0.31021 & \hspace{1cm} 1.35357\\
     vortex \hspace{2cm} & 0.28607 \hspace{1cm} &0.26017  \hspace{1cm}          &  0.30867  & \hspace{1cm} 1.33833\\
    \end{tabular}
    \caption{Complete table of average $1^{\rm st}$ Wasserstein distance for each architecture and substructure type. Here the architectures were trained on lensing images {\it without} substructure.}
    \label{tab:complete-wass}
\end{table}

\begin{figure*}[h!]
    \centering
    \includegraphics[width=\linewidth]{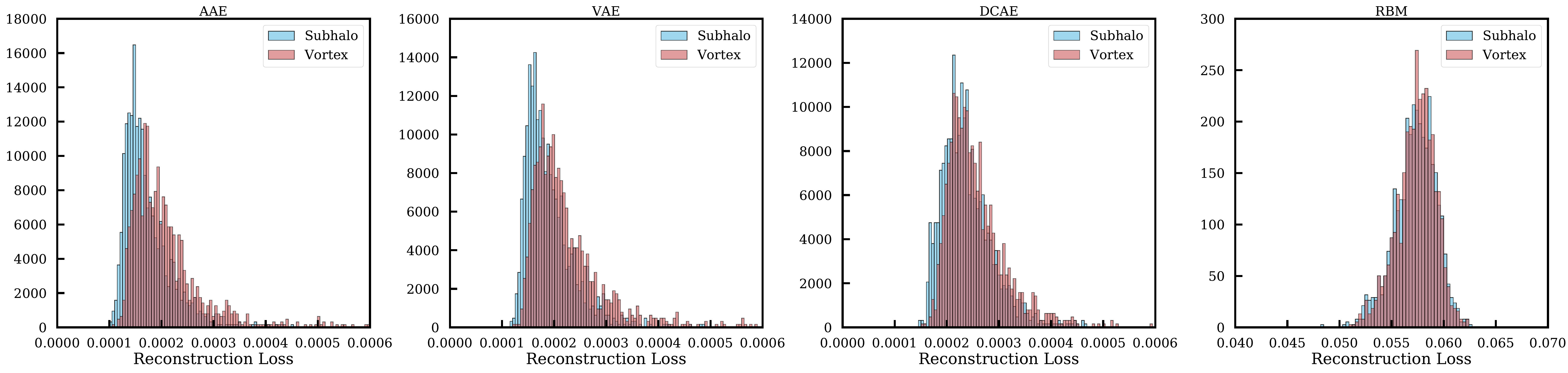}
    \caption{Reconstruction loss for unsupervised architectures for comparing subhalo and vortex performance.}
    \label{fig:recon-loss-sub-a}
\end{figure*}

\begin{figure}[!h]
    \centering
    \includegraphics[width=\linewidth]{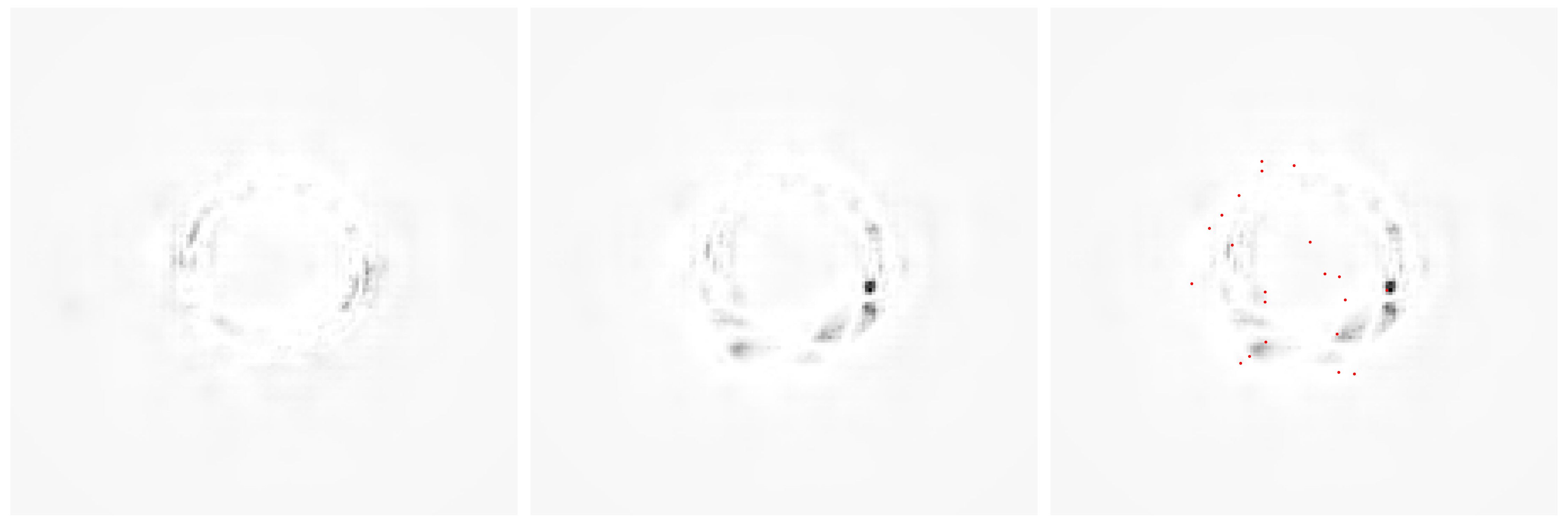}
    \caption{MSE loss for identical simulated images (save substructure) with no substructure (left), subhalo substructure (middle), and subhalo substructure with labeled subhalo position (right).}
    \label{fig:sub-mse}
\end{figure}

\clearpage
\newpage
\section{Architectures}

\begin{table}[!h]
    \footnotesize
	\centering
	\caption{Architecture of Deep Convolutional Autoencoder Model}
	\label{tab:table-DCAE}
	\begin{tabular}{cccc} 
		\hline
		\hline
		Encoder Network\\
		\hline
		\hline
		\textbf{Layer} & \textbf{Parameters} & \textbf{PyTorch Notation} & \textbf{Output Shape}\\
		\hline
Convolution  & \Longunderstack{ channels = 16 \\ kernel = 7x7 \\ stride = 3 \\ padding = 1} & Conv2d(1,16,7,3,1) &  [-1, 16, 49, 49]\\
Rectified Linear Unit (ReLU)  & - & ReLU() & [-1, 16, 49, 49]\\
Convolution  & \Longunderstack{ channels = 32 \\ kernel = 7x7 \\ stride = 3 \\ padding = 1} & Conv2d(16,32,7,3,1) & [-1, 32, 15, 15]\\
Rectified Linear Unit (ReLU)  & - & ReLU() & [-1, 32, 15, 15]\\
Convolution  & \Longunderstack{ channels = 64 \\ kernel = 7x7} & Conv2d(32,64,7) & [-1, 64, 9, 9]\\
Flatten  & - & Flatten() & [-1, 5184]\\
Fully Connected Layer & - & Linear(5184, 1000) & [-1, 1000]\\
Batch normalization & - & BatchNorm1d(1000) & [-1, 1000]\\ 
        \hline
		\hline
		Decoder Network\\
		\hline
		\hline
		\textbf{Layer} & \textbf{Parameters} & \textbf{PyTorch Notation} & \textbf{Output Shape}\\
		\hline
Fully Connected Layer & - & Linear(1000, 5184) & [-1, 5184]\\
Reshape & - & - & [-1,64,9,9]\\
Transpose Convolution  & \Longunderstack{ channels = 32 \\ kernel = 7x7} & ConvTranspose2d(64,32,7) & [-1, 32, 15, 15]\\
Rectified Linear Unit (ReLU)  & - & ReLU() & [-1, 32, 15, 15]\\
Transpose Convolution  & \Longunderstack{ channels = 16 \\ kernel = 7x7 \\ stride = 3 \\ padding = 1 \\ output padding = 2} & ConvTranspose2d(32,16,7,3,1,2) & [-1, 16, 49, 49]\\
Rectified Linear Unit (ReLU)  & - & ReLU() & [-1, 16, 49, 49]\\
Transpose Convolution  & \Longunderstack{ channels = 1 \\ kernel = 6x6 \\ stride = 3 \\ padding = 1 \\ output padding = 2} & ConvTranspose2d(16,1,6,3,1,2) & [-1, 1, 150, 150]\\
Hyperbolic tangent  & - & Tanh() & [-1, 1, 150, 150]\\
	\end{tabular}
\end{table}

\begin{table}
    \footnotesize
	\centering
	\caption{Architecture of Convolutional Variational Autoencoder Model}
	\label{tab:table-VAE}
	\begin{tabular}{cccc} 
		\hline
		\hline
		Encoder Network\\
		\hline
		\hline
		\textbf{Layer} & \textbf{Parameters} & \textbf{PyTorch Notation} & \textbf{Output Shape}\\
		\hline
Convolution  & \Longunderstack{ channels = 16 \\ kernel = 7x7 \\ stride = 3 \\ padding = 1} & Conv2d(1,16,7,3,1) &  [-1, 16, 49, 49]\\
Rectified Linear Unit (ReLU)  & - & ReLU() & [-1, 16, 49, 49]\\
Convolution  & \Longunderstack{ channels = 32 \\ kernel = 7x7 \\ stride = 3 \\ padding = 1} & Conv2d(16,32,7,3,1) & [-1, 32, 15, 15]\\
Rectified Linear Unit (ReLU)  & - & ReLU() & [-1, 32, 15, 15]\\
Convolution  & \Longunderstack{ channels = 64 \\ kernel = 7x7} & Conv2d(32,64,7) & [-1, 64, 9, 9]\\
Flatten  & - & Flatten() & [-1, 5184]\\
Fully Connected Layers & - & Linear(5184, 1000), Linear(5184, 1000) & [-1, 1000], [-1, 1000]\\ 
        \hline
		\hline
		Decoder Network\\
		\hline
		\hline
		\textbf{Layer} & \textbf{Parameters} & \textbf{PyTorch Notation} & \textbf{Output Shape}\\
		\hline
Fully Connected Layer & - & Linear(1000, 5184) & [-1, 5184]\\
Reshape & - & - & [-1,64,9,9]\\
Transpose Convolution  & \Longunderstack{ channels = 32 \\ kernel = 7x7} & ConvTranspose2d(64,32,7) & [-1, 32, 15, 15]\\
Rectified Linear Unit (ReLU)  & - & ReLU() & [-1, 32, 15, 15]\\
Transpose Convolution  & \Longunderstack{ channels = 16 \\ kernel = 7x7 \\ stride = 3 \\ padding = 1 \\ output padding = 2} & ConvTranspose2d(32,16,7,3,1,2) & [-1, 16, 49, 49]\\
Rectified Linear Unit (ReLU)  & - & ReLU() & [-1, 16, 49, 49]\\
Transpose Convolution  & \Longunderstack{ channels = 1 \\ kernel = 6x6 \\ stride = 3 \\ padding = 1 \\ output padding = 2} & ConvTranspose2d(16,1,6,3,1,2) & [-1, 1, 150, 150]\\
Hyperbolic tangent  & - & Tanh() & [-1, 1, 150, 150]\\ 
	\end{tabular}
\end{table}

\begin{table}
    \footnotesize
	\centering
	\caption{Architecture of Adversarial Autoencoder Model}
	\label{tab:table-AAE}
	\begin{tabular}{cccc} 
		\hline
		\hline
		Encoder Network\\
		\hline
		\hline
		\textbf{Layer} & \textbf{Parameters} & \textbf{PyTorch Notation} & \textbf{Output Shape}\\
		\hline
Convolution  & \Longunderstack{ channels = 16 \\ kernel = 7x7 \\ stride = 3 \\ padding = 1} & Conv2d(1,16,7,3,1) &  [-1, 16, 49, 49]\\
Rectified Linear Unit (ReLU)  & - & ReLU() & [-1, 16, 49, 49]\\
Convolution  & \Longunderstack{ channels = 32 \\ kernel = 7x7 \\ stride = 3 \\ padding = 1} & Conv2d(16,32,7,3,1) & [-1, 32, 15, 15]\\
Rectified Linear Unit (ReLU)  & - & ReLU() & [-1, 32, 15, 15]\\
Convolution  & \Longunderstack{ channels = 64 \\ kernel = 7x7} & Conv2d(32,64,7) & [-1, 64, 9, 9]\\
Flatten  & - & Flatten() & [-1, 5184]\\
Fully Connected Layer & - & Linear(5184, 1000) & [-1, 1000]\\ 
        \hline
		\hline
		Decoder Network\\
		\hline
		\hline
		\textbf{Layer} & \textbf{Parameters} & \textbf{PyTorch Notation} & \textbf{Output Shape}\\
		\hline
Fully Connected Layer & - & Linear(1000, 5184) & [-1, 5184]\\
Reshape & - & - & [-1,64,9,9]\\
Transpose Convolution  & \Longunderstack{ channels = 32 \\ kernel = 7x7} & ConvTranspose2d(64,32,7) & [-1, 32, 15, 15]\\
Rectified Linear Unit (ReLU)  & - & ReLU() & [-1, 32, 15, 15]\\
Transpose Convolution  & \Longunderstack{ channels = 16 \\ kernel = 7x7 \\ stride = 3 \\ padding = 1 \\ output padding = 2} & ConvTranspose2d(32,16,7,3,1,2) & [-1, 16, 49, 49]\\
Rectified Linear Unit (ReLU)  & - & ReLU() & [-1, 16, 49, 49]\\
Transpose Convolution  & \Longunderstack{ channels = 1 \\ kernel = 6x6 \\ stride = 3 \\ padding = 1 \\ output padding = 2} & ConvTranspose2d(16,1,6,3,1,2) & [-1, 1, 150, 150]\\
Hyperbolic tangent  & - & Tanh() & [-1, 1, 150, 150]\\ 
        \hline
		\hline
		Discriminator Network\\
		\hline
		\hline
		\textbf{Layer} & \textbf{Parameters} & \textbf{PyTorch Notation} & \textbf{Output Shape}\\
		\hline
Fully Connected Layer & - & Linear(1000, 256) & [-1, 256]\\
Rectified Linear Unit (ReLU)  & - & ReLU() & [-1, 256]\\
Fully Connected Layer & - & Linear(256, 256) & [-1, 256]\\
Rectified Linear Unit (ReLU)  & - & ReLU() & [-1, 256]\\
Fully Connected Layer & - & Linear(256, 1) & [-1, 1]\\
Sigmoid & - & Sigmoid() & [-1, 1]\\ 
	\end{tabular}
\end{table}

\end{document}